\documentclass[fleqn,usenatbib]{mnras}

 \usepackage{amsmath}
\usepackage{txfonts}
\usepackage{stfloats}   % 在导言区加入

\usepackage[T1]{fontenc}

\DeclareRobustCommand{\VAN}[3]{#2}
\let\VANthebibliography\thebibliography
\def\thebibliography{\DeclareRobustCommand{\VAN}[3]{##3}\VANthebibliography}

\usepackage{graphicx}	% Including figure files
\usepackage{amsmath}	% Advanced maths commands

\usepackage{hyperref}
\usepackage{rotating}
\usepackage{longtable}
\usepackage{multirow}
\usepackage[normalem]{ulem}
\usepackage[flushleft]{threeparttable}
\usepackage{bm}% bold math
\usepackage{makecell}
\usepackage{array}

\usepackage{gensymb}
\usepackage{ragged2e}

\usepackage{xcolor}
\title[Tightening WDM Bounds with GRBs]
{Tightening Bounds on Warm Dark Matter with High-Redshift Gamma-Ray Bursts}

\author[Wei et al.]
{Jun-Jie Wei$^{1,2}$\thanks{E-mail: jjwei@pmo.ac.cn},
Jing-Meng Hao$^{3}$,
Ding-Fang Hu$^{1,2}$,
Yang Liu$^{1}$,
Bao Wang$^{1,2}$,
Xi Kang$^{4}$\thanks{E-mail: kangxi@zju.edu.cn},
and Xue-Feng Wu$^{1,2}$\thanks{E-mail: xfwu@pmo.ac.cn}
\\
$^{1}$Purple Mountain Observatory, Chinese Academy of Sciences, Nanjing 210023, China\\
$^{2}$School of Astronomy and Space Sciences, University of Science and Technology of China, Hefei 230026, China\\
$^{3}$Department of Astronomy, School of Physics and Materials Science, Guangzhou University, Guangzhou 510006, China\\
$^{4}$Institute for Astronomy, School of Physics, Zhejiang University, Hangzhou 310027, China
}

\date{Accepted XXX. Received YYY; in original form ZZZ}

\pubyear{\the\year{}}

\begin{document}

\label{firstpage}
\pagerange{\pageref{firstpage}--\pageref{lastpage}}

\maketitle

\begin{abstract}
The cold dark matter paradigm successfully explains large-scale structure
but faces persistent tensions on small scales. Warm dark matter (WDM) with
$\mathrm{keV}$-scale particles can alleviate these issues by suppressing
small-scale structure formation. The presence of collapsed structures at
high redshifts places strong lower limits on the WDM particle mass $m_x$.
Gamma-ray bursts (GRBs) are ideal high-redshift probes due to their extreme
brightness. Using the most recent \emph{Swift} GRB data accumulated over the past
two decades, we derive robust constraints on $m_x$ by conservatively assuming
that the comoving GRB formation rate is proportional to the cosmic star formation
rate (SFR), with an additional redshift evolution parameterized as $(1+z)^\alpha$.
Applying a maximum-likelihood analysis to 118 GRBs with redshift $z<10$ and
luminosity $L\ge 4.0\times10^{52}\,\mathrm{erg\,s^{-1}}$, we obtain
$m_x \gtrsim 1.23\,\mathrm{keV}$ and $\alpha=1.57^{+1.14}_{-0.57}$ at the
95\% confidence level (CL). The no-evolution scenario ($\alpha=0$), in which
the GRB rate exactly traces the SFR without additional redshift evolution,
is excluded at the $5\sigma$ level. Adopting the best-fit value $\alpha=1.57$
as a prior tightens the lower limit on $m_x$ to $m_x \gtrsim 1.59\,\mathrm{keV}$
at the 95\% CL. These robust constraints demonstrate that GRBs are a powerful
probe of the early Universe. A better understanding of the relationship between
the GRB rate and the SFR would enable even tighter limits on WDM models.
\end{abstract}


%%%%%%%%%%%%%%%%%%%%%%%%%%%%%%%%%%%%%%%%%%%%%%%%%%

%%%%%%%%%%%%%%%%% BODY OF PAPER %%%%%%%%%%%%%%%%%%

\section{Introduction}
The cold dark matter (CDM) paradigm has been remarkably successful in explaining
the formation of large-scale structure in the Universe (e.g., \citealt{2006Natur.440.1137S,
2006PhRvD..74l3507T,2010PhR...495...33B,2020A&A...641A...6P}). On small scales
($\lesssim 1\,\mathrm{Mpc}$), however, persistent tensions between CDM predictions
and observations remain (e.g., \citealt{2012MNRAS.421.2384M,2015PNAS..11212249W,
2017ARA&A..55..343B,2017Galax...5...17D,2026NatAs..10..440V}). These include:
(i) the ``missing satellite'' problem -- CDM simulations overpredict the number
of satellite galaxies around the Milky Way (e.g.,
\citealt{1999ApJ...522...82K,1999ApJ...524L..19M,2025ApJ...988..136J}); (ii)
the ``core--cusp'' problem -- simulations predict cuspy dark matter (DM) halos,
whereas observations indicate cored profiles (e.g., \citealt{1994Natur.370..629M,
2001ApJ...552L..23D,2021Galax...9..123D}); and (iii) the ``too big to fail'' problem
-- the most massive predicted CDM subhalos are too kinematically massive to be
consistent with the observed bright Milky Way satellites
(e.g., \citealt{2011MNRAS.415L..40B,2012MNRAS.422.1203B}).

The inclusion of baryonic feedback is a widely adopted approach to addressing
the small-scale challenges of the CDM paradigm. Baryonic feedback, driven
primarily by supernova explosions and Ultraviolet (UV) background heating, can suppress
satellite galaxy formation and flatten inner halo density profiles, thereby
alleviating some of the observed discrepancies (e.g., \citealt{2000ApJ...539..517B,2008Sci...319..174M,
2011MNRAS.415.2969D,2012MNRAS.421.3464P,2014Natur.506..171P}). Nevertheless,
baryonic effects remain difficult to model accurately. They rely on simplified
sub-grid physics, and even with fine-tuning, they are insufficient to fully
reconcile observations with CDM predictions (e.g.,
\citealt{2012MNRAS.422.1203B,2022NatAs...6..897S}).

An alternative solution to the small-scale problems of CDM is to assume that
DM consists of lighter particles, with masses in the $\mathrm{keV}$
range, i.e., warm dark matter (WDM; e.g., \citealt{2001ApJ...556...93B,
2008PhRvD..78f5040K,2012NewA...17..653D,2026A&A...706A.382S}). WDM particles
retain a non-negligible velocity dispersion, leading to free-streaming and
an associated effective pressure, both of which suppress structure formation
on small scales (e.g.,
\citealt{2014MNRAS.439..300L,2019MNRAS.484.5400M,2024MNRAS.527.2835S}).
This scenario preserves the successful large-scale structure predictions of CDM
(e.g., \citealt{2003ApJ...598...49Z}) while simultaneously alleviating some
small-scale tensions (e.g., \citealt{2012MNRAS.420.2318L}), although some degree
of fine-tuning may still be required to fully match all observations (e.g.,
\citealt{2012MNRAS.424.1105M}).

The high-redshift Universe provides the most powerful testbed for the WDM
scenario. In WDM models, structure formation is exponentially suppressed
below a characteristic scale set by the free-streaming length of the WDM
particles (e.g.,
\citealt{2012MNRAS.424..684S,2013MNRAS.433.1573S}). Because structures form
hierarchically, the first galaxies are expected to reside in the smallest
DM halos. If DM were sufficiently `warm' (i.e., if the WDM particle mass
were sufficiently low), the suppression of small-scale structure would be
so severe that the high-redshift Universe would be essentially devoid of
galaxies. Consequently, the detection of a single galaxy at high redshift
can impose strong lower limits on the WDM particle mass, ruling out models
that would over-suppress structure formation too strongly at those epochs.

Current constraints on the WDM particle mass $m_x$ come from a variety of
observations. For example, the requirement that cosmological reionization
occurred at $z>5.8$ implies $m_x \gtrsim 0.75\,\mathrm{keV}$ \citep{2001ApJ...558..482B}.
Similarly, the stellar mass function and the Tully-Fisher relation require
that WDM models not be too warm, i.e., $m_x >0.75\,\mathrm{keV}$ \citep{2013ApJ...767...22K}.
Two lensed high-redshift galaxies at $z\approx10$ yield a slightly stronger
limit of $m_x >0.9\,\mathrm{keV}$ \citep{2013MNRAS.435L..53P}.
The Lyman-$\alpha$ forest flux power spectrum provides a tighter constraint,
giving $m_x \gtrsim 1.2$--$5.7\,\mathrm{keV}$
(e.g., \citealt{2008PhRvL.100d1304V,2013PhRvD..88d3502V,2017PhRvD..96b3522I,2024PhRvD.109d3511I}).
UV luminosity function (LF) measurements of high-redshift galaxies from the
Hubble Space Telescope and James Webb Space Telescope (JWST) yield a lower bound of
$m_x >1.8\,\mathrm{keV}$ at the 95\% confidence level (CL; \citealt{2026JCAP...07..010C}).
Observations of the CMB optical depth and ionization fraction place a constraint of
$1\,\mathrm{keV}\lesssim m_x \lesssim3\,\mathrm{keV}$ \citep{2016ApJ...829...29T},
while combined measurements of the galaxy LF out to $z\sim10$ and cosmic reionization
give a narrower range of $2\,\mathrm{keV}<m_x <3\,\mathrm{keV}$
\citep{2015JCAP...09..003L}. A significantly stronger lower limit of
$m_x >6.20\,\mathrm{keV}$ is obtained from the Milky Way and Andromeda
satellite populations \citep{2026ApJ..1000...88L}.  Looking ahead,
under the assumption of a CDM background, upcoming line-intensity mapping surveys
are expected to set a 95\% CL lower limit on $m_x$ of $1.10\,\mathrm{keV}$
in the optimistic scenario and $0.58\,\mathrm{keV}$ in the pessimistic scenario,
respectively \citep{2026A&A...707A.148M}.

As the most energetic explosive transients in the cosmos, gamma-ray bursts
(GRBs) can be detected up to very high redshifts, as demonstrated by the
detection of GRB 090429B at $z\sim9.4$ \citep{2011ApJ...736....7C}. GRBs
are therefore considered a powerful probe for exploring the high-redshift
Universe and small-scale structures (see
\citealt{2015JHEAp...7...35S,2015NewAR..67....1W} for reviews). In particular,
long GRBs with durations $T_{90}>2\,\mathrm{s}$ (where $T_{90}$ is the 90\%
prompt emission time interval; \citealt{1993ApJ...413L.101K}) originate from
the core collapse of massive stars (e.g., \citealt{1993ApJ...405..273W,1998ApJ...494L..45P,2006ARA&A..44..507W}),
making them excellent tracers of star formation in the early Universe (e.g.,
\citealt{1997ApJ...486L..71T,2007ApJ...671..272C,2008ApJ...683L...5Y,2009ApJ...705L.104K,
2009MNRAS.400L..10W,2012ApJ...744...95R,2013ApJ...772...42H,2013A&A...556A..90W,
2014MNRAS.439.3329W,2016JHEAp...9....1W,2020ApJS..248...21H,2024ApJ...976L..16M,
2025ApJ...988L..71W}). Importantly, it has been proposed that a lone GRB
detected at $z>10$ would be sufficient to strongly constrain WDM models
\citep{2005ApJ...623....1M}. Following this idea, \citet{2013MNRAS.432.3218D}
placed robust lower limits on the WDM particle mass $m_x$ using the \emph{Swift}
database of high-redshift long GRBs. By parameterizing the redshift evolution of
the ratio between the comoving GRB formation rate and the cosmic star formation rate
(SFR) as $(1+z)^\alpha$ to account for astrophysical uncertainties, they obtained
$m_x \gtrsim 1.6$--$1.8\,\mathrm{keV}$ at the 95\% CL after
marginalizing over a flat prior in $\alpha$.

Since the launch of the \emph{Swift} satellite in November 2004 \citep{2004ApJ...611.1005G},
the number of long GRBs with measured redshifts has increased substantially.
In this work, we update and expand the sample of \emph{Swift}/Burst Alert Telescope
(BAT)-detected long GRBs with known redshifts. Our principal goal is to use
the latest \emph{Swift} GRB data accumulated over the past
two decades to impose new, robust constraints on WDM models.
We adopt a self-consistent method within the framework of hierarchical structure
formation in a WDM-dominated Universe to construct the cosmic SFR. According to
the collapsar model, the expected redshift distribution of long GRBs can be
derived from this SFR model as a function of the WDM particle mass
$m_x$. Consequently, $m_x$ can be constrained by directly comparing the observed
redshift distribution with the expected one.

%However, all the limits discussed above suffer from a strong degeneracy between
%astrophysical (i.e., baryonic) processes and the WDM particle mass. In contrast,
%our approach is more robust because it relies solely on the shape of the redshift
%evolution of the SFR, which is directly traced by long GRBs. Notably, in WDM models
%the SFR is exponentially suppressed at high redshifts, a characteristic signature
%that cannot be reproduced by any known astrophysical uncertainties \citep{2013MNRAS.432.3218D}.
%High-redshift probes such as GRBs therefore offer a powerful means of constraining
%WDM models.

The remainder of this paper is structured as follows. In Section~\ref{sec:WDM model},
we describe hierarchical structure formation in the context of the WDM paradigm.
In Section~\ref{sec:CSFR}, we present our construction of the cosmic star formation
history and illustrate the impact of different WDM models on the derived SFR.
In Section~\ref{sec:Result}, we describe our analysis method and present the
resulting constraints obtained from the latest \emph{Swift} GRB data. Finally,
we summarize our main findings and provide a brief discussion in Section~\ref{sec:summary}.
Throughout this paper, we adopt the \emph{planck} $\Lambda$CDM cosmology with
the following parameters: $H_{0}=67.4$ $\mathrm{km\,s^{-1}\,Mpc^{-1}}$,
$\Omega_{\rm m}=0.315$, $\Omega_{\Lambda}=0.685$, $\Omega_{\rm b}=0.0493$,
and $\sigma_8=0.811$ \citep{2020A&A...641A...6P}.

\section{Hierarchical Structure Formation in the WDM Scenario}
\label{sec:WDM model}
In the hierarchical formation scenario, the Press-Schechter formalism
provides a semi-analytic estimate of the abundance of collapsed DM halos
\citep{1974ApJ...187..425P}. An improved version, the Sheth-Tormen (ST)
formalism, accounts for ellipsoidal collapse and has been shown to be
more accurate \citep{1999MNRAS.308..119S}. We therefore adopt the ST formalism
to describe the halo mass function:
\begin{equation}
f_{\rm ST}(\sigma)=A_{\rm ST}\sqrt{\frac{2a_{1}}{\pi}}\left[1+\left(\frac{\sigma^{2}}{a_{1}\delta^{2}_{c}}\right)^{p_{1}}\right]
\frac{\delta_{c}}{\sigma}\exp\left(-\frac{a_{1}\delta^{2}_{c}}{2\sigma^{2}}\right)\;,
\end{equation}
where $\sigma$ is the variance of the linear density field on the mass scale
of interest, $\delta_{c}=1.686$ is the critical linear overdensity for collapse,
and the parameters $A_{\rm ST}=0.3222$, $a_{1}=0.707$, and $p_{1}=0.3$ are chosen
to match numerical simulations \citep{1999MNRAS.308..119S}. The number density
of DM halos at a given redshift, $n_{\rm ST}(M,\,z)$, in the mass interval
$[M,\,M+\mathrm{d}M]$, is related to $f_{\rm ST}(\sigma)$ by
\begin{equation}
\frac{\mathrm{d}n_{\rm ST}(M,\,z)}{\mathrm{d}M}=\frac{\rho_{\rm m}}{M}\frac{\mathrm{d}\ln \sigma^{-1}}{\mathrm{d}M}f_{\rm ST}(\sigma)\;,
\end{equation}
where $\rho_{\rm m}$ is the mean mass density of the background Universe.
The variance of the linear density field $\sigma(M,\,z)$ is given by
\begin{equation}
\sigma^{2}(M,\,z)=\frac{D^{2}(z)}{2\pi^{2}}\int^{\infty}_{0}k^{2}P(k)W^{2}\left(k,\,M\right) \mathrm{d}k\;,
\end{equation}
where $k$ is the wavenumber, $P(k)$ is the power spectrum of density fluctuations
at $z=0$, and $W(k,\,M)$ is the window function. As shown by \cite{2013MNRAS.428.1774B}
and \cite{2014MNRAS.442.2487K}, the commonly adopted top-hat window function is
unsuitable when the power spectrum features a cutoff, as in the WDM case (see below);
instead, a sharp-$k$ filter is employed. We refer the reader to \cite{2014MNRAS.442.2487K}
for a detailed discussion (see also \citealt{2020MNRAS.491.2520K}). The sharp-$k$ filter
is given by
\begin{equation}
    W(k,\,M) = \left\lbrace \begin{array}{ll} 1; ~~~~~k \leq k_{s}(R) \\
    0; ~~~~~k > k_{s}(R )\;, \\
\end{array} \right.
\end{equation}
where $R=(3M/4\pi\rho_{\rm m})^{1/3}$ is the radius of a halo of mass $M$, and
$k_{s}=2.5/R$ \citep{2013MNRAS.428.1774B}. The redshift dependence of $\sigma(M,\,z)$
enters only through the growth factor $D(z)$, which can be expressed as
$D(z)=g(z)/[g(0)(1+z)]$, where $g(z)$ is approximated by
\begin{equation}
g(z)\approx\frac{350\,\Omega_{\rm m}(z)}{2+209\,\Omega_{\rm m}(z)-\Omega_{\rm m}^{2}(z)+140\,\Omega_{\rm m}^{4/7}(z)}\;,
\end{equation}
with $\Omega_{\rm m}(z)=\Omega_{\rm m}(1+z)^3 / [\Omega_{\rm m}(1+z)^3 + \Omega_{\Lambda}]$.
Thus, $\sigma(M,\,z)=\sigma(M,\,0)D(z)$.

In WDM models, the amplitude of linear perturbations is damped on scales smaller
than the free-streaming length. The smaller the WDM particle mass, the larger
the characteristic scale below which the power spectrum is suppressed. The
comoving free-streaming scale is given by
(see, e.g., \citealt{1996ApJ...458....1C,2001ApJ...556...93B,2005PhRvD..71f3534V})
\begin{equation}
\lambda_{\rm fs}\approx0.11\left(\frac{\Omega_xh^2}{0.15}\right)^{1/3}\left(\frac{m_x}{\mathrm{keV}}\right)^{-4/3}\,\mathrm{Mpc} \,,
\end{equation}
where $\Omega_x$ is the density parameter of the WDM particle (i.e., its energy density
relative to the critical density), $h\equiv H_0/(\mathrm{100\,km\,s^{-1}\,Mpc^{-1}})$
is the dimensionless Hubble constant, and $m_x$ is the WDM particle mass.
The modification to the primordial power spectrum is obtained by multiplying
the CDM power spectrum $P_{\rm CDM}(k)$ by a transfer function \citep{2001ApJ...556...93B}:
\begin{equation}
P_{\rm WDM}(k)=A_{P}P_{\rm CDM}(k)T^{2}(k)\,,
\end{equation}
where $A_{P}$ is the normalization factor chosen to reproduce the required
$\sigma_8=0.811$ \citep{2020A&A...641A...6P} when integrated under a sharp-$k$
filter with radius $8\,h^{-1}\,\mathrm{Mpc}$. The WDM transfer function is
then given by \citep{2005PhRvD..71f3534V}
\begin{equation}
T(k)=\left[1+\left(\epsilon k\right)^{2\mu}\right]^{-5/\mu}\,,
\end{equation}
where $\mu=1.12$ and the parameter $\epsilon$ is related to the WDM particle mass
$m_x$ via
\begin{equation}
\epsilon=0.049\left(\frac{\Omega_x}{0.25}\right)^{0.11}\left(\frac{m_x}{\mathrm{keV}}\right)^{-1.11}\left(\frac{h}{0.7}\right)^{1.22}\,h^{-1}\,\mathrm{Mpc}\,.
\end{equation}
We compute the CDM power spectrum $P_{\rm CDM}(k)$ using the CAMB code.\footnote{\url{https://camb.info/}}

In WDM models, the growth of density perturbations is further delayed by
the residual velocity dispersion of WDM particles, which introduces an
effective pressure. This effective pressure has been shown to be as
important as the power-spectrum cutoff in shaping the WDM mass function
(e.g., \citealt{2001ApJ...558..482B,2005ApJ...623....1M}). Using spherical
hydrodynamical simulations and an analogy with gas pressure, \cite{2001ApJ...558..482B}
defined an effective WDM Jeans mass---the mass scale below which gravitational
collapse is significantly suppressed by this pressure. Following
\cite{2013MNRAS.432.3218D}, we denote this characteristic mass as
\begin{equation}
M_{\rm WDM}\approx1.8\times10^{10}\left(\frac{\Omega_x h^2}{0.15}\right)^{1/2}\left(\frac{m_x}{\mathrm{keV}}\right)^{-4}\,\mathrm{M_{\odot}}\,.
\end{equation}

We assume that baryons trace the DM distribution in an unbiased manner,
i.e., the baryonic density is proportional to the DM density.\footnote{
The quantity $M_{\rm WDM}(m_x)$ is a purely DM threshold, determined
solely by the WDM power-spectrum cutoff \citep{2001ApJ...558..482B},
and is therefore independent of baryonic processes. The validity of
the ``baryons trace DM'' (BTD) approximation was systematically examined
by \cite{2020ApJ...900...30P}, who found that it overestimates the halo mass
function by only about $5\%$. Given that WDM free-streaming suppresses
the abundance of low-mass halos by several orders of magnitude, this
$5\%$ bias is entirely negligible for our analysis. Therefore, our derived
limits on $M_{\rm WDM}(m_x)$ and $m_x$ remain robust against the BTD assumption.}
Star formation is assumed to occur only in halos that are sufficiently dense;
this condition is parametrized by a threshold halo mass $M_{\rm min}$.
The fraction of baryons residing in collapsed halos at redshift $z$ is then given by
\begin{equation}
\label{eq:fb}
f_{\rm b}(z)=\frac{\int^{\infty}_{M_{\rm min}}n_{\rm ST}(M,\,z)M\,\mathrm{d}M}
{\rho_{\rm m}} \;,
\end{equation}
where $M_{\rm min}$ is the minimum halo mass capable of hosting star formation,
estimated as
\begin{equation}
M_{\rm min}=\max\left[M_{\rm gal}(z),\,M_{\rm WDM}(m_x)\right] \;.
\end{equation}
Here $M_{\rm gal}(z)$ denotes the evolving minimum halo mass that
accounts for both the gas cooling criterion and radiative feedback
from the ionizing UV background during inhomogeneous reionization
\citep{2013MNRAS.432L..51S}.\footnote{Applying the $M_{\rm gal}(z)$
prescription of \cite{2013MNRAS.432L..51S} to WDM models is not strictly
self-consistent, as the UV background feedback they model should be weaker
in the WDM case. This inconsistency is, however, smaller than other
astrophysical uncertainties. Crucially, our conclusions are driven by
models for which $M_{\rm WDM}>M_{\rm gal}$ at high redshifts, rendering
our results insensitive to the specific choice of $M_{\rm gal}$.}
It is expected to evolve from $M_{\rm cool}$ at high redshifts to $M_0$
at low redshifts, with a transition occurring at $z\sim z_{\rm re}-\Delta z_{\rm sc}$
(where $z_{\rm re}$ is the redshift of homogeneous reionization and
$\Delta z_{\rm sc}$ corresponds to the sound-crossing time scale),
over a duration roughly scaling with the reionization width $\Delta z_{\rm re}$.
Following \cite{2013MNRAS.432L..51S}, $M_{\rm gal}(z)$ is given by
\begin{equation}
M_{\rm gal}(z)=M_{\rm cool}\left(\frac{M_0}{M_{\rm cool}}\right)^{y}\;,
\end{equation}
where $y(z)=\left(1+\exp\left[\frac{z-\left(z_{\rm re}-\Delta z_{\rm sc}\right)}{\Delta z_{\rm re}}\right]\right)^{-1}$
is the transition function, and $M_{\rm cool}$ is the cooling threshold
required for gas to efficiently cool, collapse, and form stars inside
a halo with a constant virial temperature $T_{\rm vir}\sim10^4\,\mathrm{K}$.
This threshold is equivalent to a redshift-dependent halo mass \citep{2001PhR...349..125B}:
\begin{equation}
\begin{aligned}
M_{\rm cool}(z)=&10^8 h^{-1}\left(\frac{\bar{w}}{0.6}\right)^{-3/2}\left(\frac{\Omega_{\rm m}}{\Omega_{\rm m}(z)}\frac{\Delta_{c}}{18\pi^2}\right)^{-1/2}\\
&\times\left(\frac{T_{\rm vir}}{1.98\times10^4\,\mathrm{K}}\right)^{3/2}\left(\frac{1+z}{10}\right)^{-3/2}\,\mathrm{M_{\odot}}\;,
\end{aligned}
\end{equation}
where $\bar{w}=0.6$ is the mean molecular  weight, $\Delta_{c}=18\pi^2+82d-39d^2$
with $d=\Omega_{\rm m}(z)-1$.
Following \cite{2013MNRAS.432L..51S}, we take $M_0 (z)$ to correspond
to $M_{\rm cool}(T_{\rm vir})$ with a fixed virial temperature
$T_{\rm vir}=5\times10^4\,\mathrm{K}$, and adopt the standard parameters
$z_{\rm re}=9.3$, $\Delta z_{\rm re}=1.0$, and $\Delta z_{\rm sc}=2.0$.
Note that for models with $m_{x}\lesssim2\,\mathrm{keV}$, we have
$M_{\rm WDM}(m_x)>M_{\rm gal}(z)$ at $z\gtrsim6$; consequently, $M_{\rm min}$
becomes insensitive to the exact value of $M_{\rm gal}$ at sufficiently
high redshifts. For models with $m_{x}=4\,\mathrm{keV}$, the condition
$M_{\rm WDM}(m_x)>M_{\rm gal}(z)$ holds only at $z\gtrsim13$. For heavier
WDM particles, the free-streaming scale and hence $M_{\rm WDM}$ decreases.
As a result, when $M_{\rm WDM}$ falls below the typical $M_{\rm gal}(z)$
relevant at high redshifts, the suppression of galaxy formation is no longer
dominated by WDM effects; even in a pure CDM Universe, galaxy formation is
suppressed below halo masses comparable to $M_{\rm WDM}$. Therefore,
constraints on WDM masses larger than $\sim4\,\mathrm{keV}$ cannot be
robustly obtained from the present analysis. Tighter constraints on heavier
$m_{x}$ would require observations at significantly higher redshifts.

With the baryonic fraction given by Equation~(\ref{eq:fb}), the baryonic accretion
rate $a_{\rm b}(t)$, which describes the process of structure formation,
can be calculated as \citep{2006ApJ...647..773D}
\begin{equation}
\label{eq:ab}
a_{\rm b}(t)=\Omega_{\rm b}\rho_{\rm c}\left(\frac{\mathrm{d}t}{\mathrm{d}z}\right)^{-1}
\left|\frac{\mathrm{d} f_{\rm b}(z)}{\mathrm{d}z}\right| \;,
\end{equation}
where $\rho_{\rm c}=3H^{2}_{0}/8\pi G$ is the critical density of the Universe.
The age of the Universe is related to redshift by
\begin{equation}
\frac{\mathrm{d}t}{\mathrm{d}z}=\frac{9.78\,h^{-1}\,\mathrm{Gyr}}{\left(1+z\right)\sqrt{\Omega_\Lambda +\Omega_{\rm m}\left(1+z\right)^3}}\;.
\end{equation}

In the next section, we discuss how to obtain the cosmic SFR in WDM models
and examine the effect of the WDM particle mass on the SFR.

\section{The Cosmic Star Formation}
\label{sec:CSFR}
Generally, the star formation history of a galactic-like system is governed by
the balance of baryons between two large reservoirs in the Universe \citep{2006ApJ...647..773D}.
The first reservoir is associated with collapsed structures (hereafter ``structures'')
and comprises two subreservoirs: the interstellar medium (ISM) and the stars
(including their remnants). The second reservoir corresponds to the intergalactic
medium (IGM) between the collapsed structures. Matter is exchanged among these
three baryonic (sub)reservoirs (IGM, ISM, and stars) through four physical processes:
(a) accretion of baryons from the IGM onto forming structures, $a_{\rm b}(t)$
(Equation~\ref{eq:ab}); (b) conversion of ISM gas into stars, $\dot{\rho}_{\star}(t)$;
(c) ejection of mass from stars back into the ISM via stellar winds and supernovae,
$M_{\rm ej}(t)$; and (d) outflow of baryons from structures into the IGM, $o(t)$.
Therefore, the evolution of the total gas density $\rho_{\rm g}(t)$ within DM halos,
which governs the star formation history, is described by
\citep{2006ApJ...647..773D,2016ApJ...829...29T}
\begin{equation}
\label{eq:gas}
\dot{\rho}_{\rm g}=-\frac{\mathrm{d}^2M_{\star}}{\mathrm{d}V\mathrm{d}t}
+\frac{\mathrm{d}^2M_{\rm ej}}{\mathrm{d}V\mathrm{d}t}+a_{\rm b}(t)-o(t)\;,
\end{equation}
where the first term on the right-hand side is the SFR, i.e., the rate at which
gas is converted into stars; the second term is the mass ejection rate from stars;
the third term is the baryonic accretion rate; and the fourth term is the baryonic outflow rate.

We calculate the SFRs of both Population II/I (Pop II/I) and Population III (Pop III) stars.
For simplicity, we assume that each SFR $\dot{\rho}_{\star}(t)$ follows
the Schmidt law \citep{1959ApJ...129..243S,1963ApJ...137..758S}, i.e.,
the SFR is proportional to the gas density $\rho_{\rm g}(t)$ in star-forming
structures. Additionally, we adopt an exponentially declining SFR for
Pop II/I stars, which has been shown to reproduce the observational data well
\citep{2006ApJ...647..773D}. For Pop II/I stars, we therefore take
\begin{equation}
\label{eq:SFRII}
\dot{\rho}_{\star,\mathrm{II/I}}(t)=f_1\frac{\rho_{\rm g}(t)}{\tau_{\rm 1}}e^{-\left(t-t_{\rm init}\right)/\tau_{1}}\left(1-e^{-Z_{\rm IGM}/Z_{\rm crit}}\right)\,,
\end{equation}
where $t_{\rm init}$ is the age of the Universe at the initial redshift
$z_{\rm init}$, and $\tau_{1}$ is the characteristic timescale for star formation.
We assume that star formation begins at $z_{\rm init}=30$, and we treat the factor
$f_1\exp[-\left(t-t_{\rm init}\right)/\tau_{1}]$ as the star formation efficiency
of Pop II/I stars, which is calibrated using the observed SFR at low redshifts.
The last term, $[1-\exp(-Z_{\rm IGM}/Z_{\rm crit})]$, accounts for the fraction
of gas that has been enriched by metals from outflowing structures, where
\begin{equation}
Z_{\rm IGM}(t)=\frac{\int_{t}^{t_{\rm init}}o(t)\,\mathrm{d}t}{\Omega_{\rm b}\rho_{\rm c}-\int_{t}^{t_{\rm init}}a_{\rm b}(t)\,\mathrm{d}t+\int_{t}^{t_{\rm init}}o(t)\,\mathrm{d}t}
\end{equation}
is the metal enrichment history of the IGM, and $Z_{\rm crit}=10^{-4}\,Z_{\odot}$
is the critical metallicity. For Pop III stars, we assume a transition from Pop III
to the normal formation mode at a critical metallicity $Z_{\rm crit}$, and parametrize
the SFR as \citep{2006ApJ...647..773D}
\begin{equation}
\label{eq:SFRIII}
\dot{\rho}_{\star,\mathrm{III}}(t)=f_2\frac{\rho_{\rm g}(t)}{\tau_{\rm 2}}e^{-Z_{\rm IGM}/Z_{\rm crit}}\,.
\end{equation}
We adopt a star formation efficiency $f_2=4.5\%$ and a characteristic timescale
$\tau_2=0.1\,\mathrm{Gyr}$ \citep{2004ApJ...617..693D}. Thus, the total rate
at which mass is converted into stars is given by the sum of the SFRs of
Pop II/I and Pop III stars:
\begin{equation}
\label{eq:Mstar}
\frac{\mathrm{d}^2M_{\star}}{\mathrm{d}V\mathrm{d}t}=
\dot{\rho}_{\star,\mathrm{II/I}}(t)+\dot{\rho}_{\star,\mathrm{III}}(t)\;.
\end{equation}

We adopt a Salpeter form for the initial mass function (IMF; \citealt{1955ApJ...121..161S})
for both Pop II/I and Pop III stars:
\begin{equation}
\varphi(m)\propto m^{-2.35}\;.
\end{equation}
The IMF is normalized such that
$\int_{m_{\rm inf}}^{m_{\rm sup}} m\varphi(m)\,\mathrm{d}m=1$,
with the integration mass range set to $m_{\rm inf}=0.1\,\mathrm{M_{\odot}}$ and
$m_{\rm sup}=100\,\mathrm{M_{\odot}}$ for Pop II/I stars, and
$m'_{\rm inf}=100\,\mathrm{M_{\odot}}$ and $m'_{\rm sup}=500\,\mathrm{M_{\odot}}$
for Pop III stars.

The mass ejected from stars back into the ISM via stellar winds and supernovae is given by
\begin{equation}
\label{eq:Mej}
\begin{aligned}
\frac{\mathrm{d}^2M_{\rm ej}}{\mathrm{d}V\mathrm{d}t}=&
\int_{m_d(t)}^{m_{\rm sup}}\left(m-m_{\rm r}\right)\varphi(m)\dot{\rho}_{\star,\mathrm{II/I}}\left(t-\tau_{m}\right)\,\mathrm{d}m \,+\\
&\int_{m'_d(t)}^{m'_{\rm sup}}\left(m'-m'_{\rm r}\right)\varphi(m')\dot{\rho}_{\star,\mathrm{III}}\left(t-\tau_{m'}\right)\,\mathrm{d}m' \;,
\end{aligned}
\end{equation}
where $\tau_{m}$ is the lifetime of a star of mass $m$, $m_d(t)$ corresponds to
the mass of stars that die at cosmic time $t$, $m_{\rm r}$ is the mass
of the stellar remnant, and $\dot{\rho}_{\star}(t-\tau_{m})$ is the SFR evaluated at
the retarded time $t-\tau_{m}$. The same definitions apply to the primed quantities
for Pop III stars. To obtain $m_d(t)$ or $\tau_{m}$, we adopt the mass--lifetime relation
\citep{1986FCPh...11....1S,1997ApJ...487..704C}:
\begin{equation}
\log_{10}\left(\frac{\tau_{m}}{\mathrm{yr}}\right)=10.0-3.6\log_{10}\left(\frac{m_d}{\mathrm{M_{\odot}}}\right)+\left[\log_{10}\left(\frac{m_d}{\mathrm{M_{\odot}}}\right)\right]^2 \;.
\end{equation}
The mass of the stellar remnant $m_{\rm r}$ depends on the progenitor mass.
We adopt the following prescription for $m_{\rm r}$ \citep{2010MNRAS.401.1924P}:
\begin{enumerate}
  \item Stars with $m<1\,\mathrm{M_{\odot}}$ have lifetimes longer than
  the Hubble time and thus do not contribute to the ejected mass $M_{\rm ej}$.
  \item Stars with $1\,\mathrm{M_{\odot}}\leq m \leq 8\,\mathrm{M_{\odot}}$
  evolve into carbon-oxygen white dwarfs, with $m_{\rm r}=0.1156\,m
  +0.4551$.
  \item Stars with $8\,\mathrm{M_{\odot}}< m \leq 10\,\mathrm{M_{\odot}}$
  become oxygen-neon-magnesium white dwarfs, with $m_{\rm r}=1.35\,\mathrm{M_{\odot}}$.
  \item Stars with $10\,\mathrm{M_{\odot}}< m < 25\,\mathrm{M_{\odot}}$ explode as
  supernovae and leave neutron stars, with $m_{\rm r}=1.4\,\mathrm{M_{\odot}}$.
  \item Stars with $25\,\mathrm{M_{\odot}}\leq m < 140\,\mathrm{M_{\odot}}$
  form black holes. In this case, $m_{\rm r}=m_{\rm He}$, where $m_{\rm He}$ is the mass
  of the helium core before collapse. Following \cite{2002ApJ...567..532H},
  $m_{\rm He}=\frac{13}{24}(m-20\,\mathrm{M_{\odot}})$.
  \item Stars with $140\,\mathrm{M_{\odot}}\leq m < 260\,\mathrm{M_{\odot}}$
  explode as pair-instability supernova \citep{2002ApJ...567..532H} and leave no remnants.
  \item Stars with $260\,\mathrm{M_{\odot}}\leq m \leq 500\,\mathrm{M_{\odot}}$
  collapse directly into black holes, with no mass loss.
\end{enumerate}

The outflow of baryons from structures into the IGM is described by
\begin{equation}
\label{eq:outflow}
\begin{aligned}
o(t)=&\frac{2\varepsilon}{v^2_{\rm sec}(z)}
\int_{\max[8\,\mathrm{M_{\odot}},\,m_d(t)]}^{m_{\rm sup}}\varphi(m)\dot{\rho}_{\star,\mathrm{II/I}}\left(t-\tau_{m}\right)E_{\rm kin}(m)\,\mathrm{d}m\,+\\
&\frac{2\varepsilon}{v^2_{\rm sec}(z)}\int_{\max[100\,\mathrm{M_{\odot}},\,m'_d(t)]}^{m'_{\rm sup}}\varphi(m')\dot{\rho}_{\star,\mathrm{III}}\left(t-\tau_{m'}\right)E'_{\rm kin}(m')\,\mathrm{d}m' \;,
\end{aligned}
\end{equation}
where $E_{\rm kin}(m)$ is the kinetic energy from the explosion of a star of mass $m$,
$\varepsilon=10^{-3}$ is the efficiency with which this energy is converted into
outflow power, and $v^2_{\rm sec}(z)$ is the mean-square escape velocity of structures
at redshift $z$ (e.g., \citealt{1997ApJ...476..521S}). Following \cite{2016ApJ...829...29T},
we set $E_{\rm kin}=10^{51}\,\mathrm{erg}$ for Pop II/I stars, and
$E'_{\rm kin}=10^{52}\,\mathrm{erg}$ for Pop III stars. The escape velocity is derived by
averaging the gravitational potential over the halo mass distribution at redshift $z$:
\begin{equation}
v^2_{\rm sec}(z)=\frac{\int^{\infty}_{M_{\rm min}}n_{\rm ST}(M,\,z)M\left(2GM/R\right)\,\mathrm{d}M}
{\int^{\infty}_{M_{\rm min}}n_{\rm ST}(M,\,z)M\,\mathrm{d}M} \;.
\end{equation}

\begin{figure}
%\vskip-0.1in
\centerline{\includegraphics[angle=0,scale=0.5]{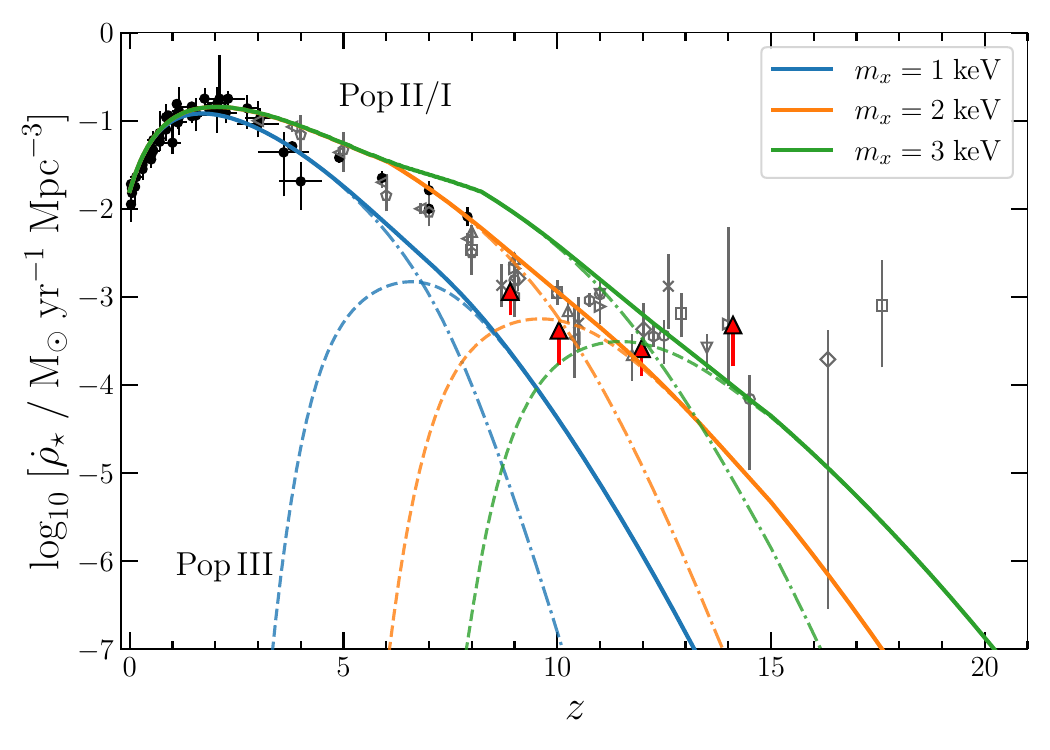}}
\caption{Redshift evolution of the cosmic SFR. The solid lines show the predicted total SFR
(including contributions from both Pop II/I and Pop III stars) for different WDM particle
masses: $m_x=1\,\mathrm{keV}$ (blue), $2\,\mathrm{keV}$ (orange), and $3\,\mathrm{keV}$
(green). The corresponding dot-dashed and dashed lines indicate the SFRs of Pop II/I and
Pop III stars, respectively. The observed SFR data include
measurements from \citet{2014ARA&A..52..415M} (black dots), as well as other high-redshift
measurements from pre-JWST photometric observations (gray open points; see
\citealt{2025ApJ...980..138H} and references therein) and spectroscopic lower limits from
JWST UV data (red triangles; \citealt{2024ApJ...960...56H,2025ApJ...980..138H}).}
\label{fig1}
\end{figure}

Finally, we derive the evolution of the total gas density $\rho_{\rm g}(t)$
at each cosmic time $t$ (or equivalently redshift $z$) by combining
Equations~(\ref{eq:ab}), (\ref{eq:Mstar}), (\ref{eq:Mej}), and (\ref{eq:outflow}) with
Equation~(\ref{eq:gas}). Once $\rho_{\rm g}(t)$ is obtained, the cosmic SFRs
$\dot{\rho}_{\star,\mathrm{II/I}}(t)$ and $\dot{\rho}_{\star,\mathrm{III}}(t)$ for
Pop II/I and Pop III stars follow directly from Equations~(\ref{eq:SFRII}) and
(\ref{eq:SFRIII}), respectively. For Pop II/I stars, we find that the model-predicted SFR,
with a star formation timescale $\tau_{1}=\mathrm{3.5\,Gyr}$, reproduces the observed SFR
at redshifts $z\lesssim4$ well. The normalization coefficient $f_1$ is determined by
requiring the total SFR (Equation~\ref{eq:Mstar}) to match
$\dot{\rho}_{\star}=0.016\,\mathrm{M_{\odot}\,yr^{-1}\,Mpc^{-3}}$ at $z=0$.

Figure~\ref{fig1} presents the total SFR obtained from our self-consistent model
for different WDM particle masses: $m_x=1\,\mathrm{keV}$ (blue solid line),
$2\,\mathrm{keV}$ (orange solid line), and $3\,\mathrm{keV}$ (green solid line).
The dot-dashed and dashed lines correspond to the SFRs of Pop II/I and Pop III stars,
respectively. The observed SFR data include measurements
from \citet{2014ARA&A..52..415M} (black dots), along with other high-redshift estimates
from pre-JWST photometric observations (gray open points; see \citealt{2025ApJ...980..138H}
and references therein) and spectroscopic lower limits from JWST UV data (red triangles;
\citealt{2024ApJ...960...56H, 2025ApJ...980..138H}). The figure clearly demonstrates
that the model predictions are in good agreement with observations at $z\lesssim4$.
However, at high redshifts, the predicted SFR varies significantly and is
highly sensitive to $m_x$.

As shown in Figure~\ref{fig1}, recent JWST observations indicate a higher SFR
at $z\gtrsim10$ than the extrapolation of pre-JWST results would suggest (e.g.,
\citealt{2023ApJS..265....5H,2024ApJ...960...56H, 2025ApJ...980..138H}). If
this excess genuinely reflects a higher star formation efficiency at high redshifts,
then a lighter WDM particle could still suppress early structure formation
while producing sufficient star formation to match the observed GRB counts,
potentially weakening the derived WDM limits. In principle, a more detailed
modeling of the redshift-dependent star formation efficiency would be valuable.
However, given the complexity of the physical processes involved and their poor
empirical constraints at high redshifts, we adopt a conservative approach in
our analysis (see below). Specifically, we parametrize the redshift evolution
of the ratio between the cosmic GRB rate and the SFR as $(1+z)^\alpha$.
This parameterization is conservative in the sense that it can effectively
absorb both the suppression of early structure formation in WDM models and
the potential evolution of the star formation efficiency inferred from JWST data.
As a result, our constraints on $m_x$ are not biased toward overly strong limits,
and our conclusions remain robust even in the presence of such astrophysical
uncertainties. Nevertheless, we acknowledge that future work incorporating
improved physical models of high-redshift star formation, informed by ongoing
JWST observations, will further strengthen the reliability of these constraints.

%however, at $z>3$, they are generally higher than most UV-based determinations.
%UV-based SFR estimates may be incomplete, potentially underestimating embedded
%star formation by up to an order of magnitude
%\citep{2012ARA&A..50..531K,2014ARA&A..52..415M,2016MNRAS.461.1100R}.
%Although early galaxies are predicted to have lower dust content than those
%at $z<3$ (e.g., \citealt{2015Natur.522..455C}), key questions regarding
%the IR luminosities of high-redshift galaxies
%\citep{2016ApJ...833...72B,2016ApJ...820...98L} and the origin of dust grains
%\citep{2015MNRAS.451L..70M,2016A&A...587A.157B,2016MNRAS.463L.112F} remain
%unresolved. Thus, dust extinction might already be significant at $z\sim7$,
%potentially hiding dusty, UV-faint galaxies in the early Universe
%\citep{2013MNRAS.429.2718S,2016MNRAS.462.3130M,2016ApJ...823..128M}.

\section{Constraining WDM with High-Redshift GRBs}
\label{sec:Result}
\subsection{The Sample}
\cite{2021MNRAS.508...52L} compiled a sample of \emph{Swift} long GRBs, comprising
424 GRBs with redshift measurements up to 2019 November. Note that the redshift of
GRB 080320 ($z=7.0$) presented in Table~A1 of \cite{2021MNRAS.508...52L} is not
reliable; therefore, we discard this burst. We extend this sample
by including additional \emph{Swift}-detected bursts that have measured redshifts
and durations satisfying $T_{90}>\mathrm{2\,s}$. Our final sample contains 496 GRBs
up to 2026 January, of which 73 are newly added. These 73 long GRBs with known
redshifts are listed in Table~\ref{table:A1}, which provides for each burst its
name, the 1-second peak photon flux $P$ in the observer-frame \emph{Swift}/BAT
energy band (15--150 $\mathrm{keV}$), and the redshift.\footnote{For GRB 210905A,
we provide the 1-second peak energy flux (in units of $\mathrm{erg\,cm^{-2}\,s^{-1}}$)
in the 15--1500 keV energy range, as reported by \cite{2022A&A...665A.125R}.}

\begin{figure}
%\vskip-0.1in
\centerline{\includegraphics[angle=0,scale=0.5]{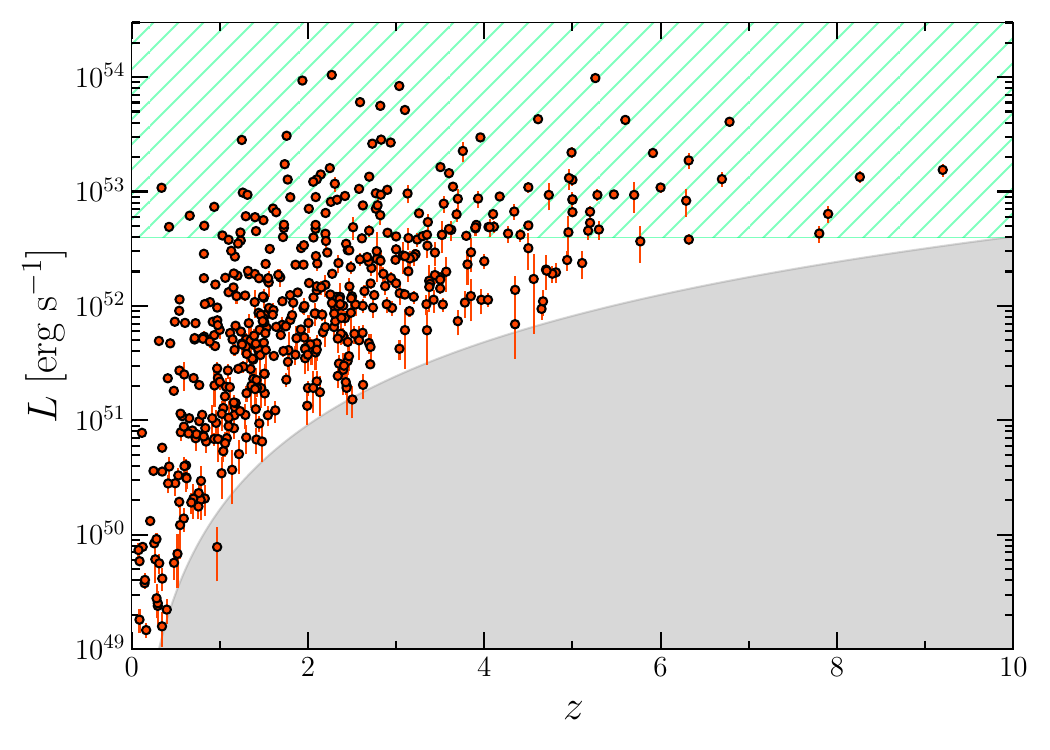}}
\caption{Luminosity--redshift distribution of 496 \emph{Swift} GRBs.
The gray shaded region approximates the luminosity threshold
(Equation~\ref{eq:Llimit}), and the green diagonal region marks
the luminosity cut $L\ge 4.0\times10^{52}\,\mathrm{erg\,s^{-1}}$
that defines our subsample.}
\label{fig2}
\end{figure}

We calculate the isotropic peak luminosity in the rest-frame energy range
1--$10^{4}$ keV for each GRB using the peak photon flux $P$:
\begin{equation}
L=4\pi D_L^2 P \frac{\int^{10^4/(1+z)\;{\rm keV}}_{1/(1+z)\;{\rm keV}} E\,N(E){\rm d}E}{\int^{150\;{\rm keV}}_{15\;{\rm keV}} N(E){\rm d}E}\;,
\end{equation}
where $D_L$ is the luminosity distance and $N(E)$ is the GRB photon spectrum.
Due to the narrow energy bandpass of the BAT, most \emph{Swift} GRB spectra
can only be fitted with a simple power law, which does not allow a direct
measurement of the broad GRB spectrum or of the spectral peak energy $E_p$.
For the 73 newly added GRBs, we therefore adopt a typical Band function spectrum
with low- and high-energy spectral indices $-1$ and $-2.3$, respectively, and
a peak energy $E_{p}=250\,\mathrm{keV}$ to calculate $L$
\citep{1993ApJ...413..281B,2000ApJS..126...19P,2006ApJS..166..298K}.
The luminosity--redshift distribution of the 496 GRBs in our final sample
is shown in Figure~\ref{fig2}. Notably, ten of these bursts lie at high redshifts
($z\ge6$).

\subsection{Analysis method}
\label{subsec:MLE}
The collapsar scenario implies that the birth of each long GRB directly marks
the demise of a short-lived massive star (e.g., \citealt{1997ApJ...486L..71T,
2000MNRAS.312L..35B,2000ApJ...536....1L,2022ApJ...932...10G}). Therefore, the
comoving GRB formation rate $\dot{\rho}_{\mathrm{GRB}}(z)$ is expected to
trace the cosmic SFR $\dot{\rho}_{\star}(z)$. A common parameterization of
this relation is
\begin{equation}\label{eq:L-sigma}
\dot{\rho}_{\mathrm{GRB}}(z)=\eta_{0}\left(1+z\right)^{\alpha}\dot{\rho}_{\star}(z)\;,
\end{equation}
where $\eta_{0}$ (in units of $\mathrm{M_{\odot}^{-1}}$) represents the GRB
formation efficiency per unit stellar mass, and $\alpha$ quantifies the
redshift dependence of this efficiency. Several mechanisms have been proposed
to account for the evolutionary trend captured by $\alpha$, including
cosmic metallicity evolution \citep{2006ApJ...638L..63L,2008MNRAS.388.1487L,
2010MNRAS.407.1972C,2014MNRAS.439.3329W}, an evolving GRB LF
\citep{2009MNRAS.396..299S,2012ApJ...749...68S,2019MNRAS.488.4607L,
2021MNRAS.508...52L,2024ApJ...976..170Q,2025ApJ...982..148Q}, an evolving stellar IMF
\citep{2009ChA&A..33..151X,2011ApJ...727L..34W}, an evolving stellar-mass
function of GRB host galaxies \citep{2022ApJ...938..129L}, and potential selection
effects \citep{2008MNRAS.386..111C,2013MNRAS.432.2141C,2012ApJ...745..168L}.
Nevertheless, the exact nature and degree of such an evolutionary trend remain
unresolved. For simplicity, we adopt a phenomenological parameterization that
describes the overall evolution of the GRB rate as $(1+z)^{\alpha}$ and
conservatively treat $\alpha$ as a free parameter.\footnote{
If the evolutionary trend described by $\alpha$ arises primarily from
cosmic metallicity evolution, then treating $\alpha$ as an independent free parameter
is physically inconsistent, since metallicity evolution is itself delay-coupled to
the suppressed structure formation in WDM models. Nevertheless, we emphasize that
our treatment is intentionally conservative. By allowing the GRB-to-SFR ratio to
evolve as a free parameter $(1+z)^{\alpha}$, we effectively absorb both the suppression
of early structure formation in WDM and a wide range of possible astrophysical
uncertainties, including those related to metallicity, feedback, and star formation
efficiency. This parameterization yields more conservative constraints on $m_x$,
thereby avoiding any overstatement of the data's discriminatory power.}
The cosmic SFR $\dot{\rho}_{\star}(z)$ is derived from a hierarchical structure formation
model that depends on a free parameter $m_x$ (the WDM particle mass), as detailed
in Section~\ref{sec:CSFR}.

When considering an instrument with a flux threshold $F_{\rm lim}$,
the expected number of GRBs with luminosity $L$ exceeding a redshift-dependent
threshold $L_{\rm lim}(z)$ and with redshift $z\in [0,\,z_{\rm max}]$
is given by
\begin{equation}
\label{eq:Number}
\begin{aligned}
N_{\rm exp}=&\Delta T\frac{\Delta\Omega}{4\pi}
\int_{0}^{z_{\rm max}}F(z)\left\langle f_{\rm beam}\right\rangle\frac{\dot{\rho}_{\rm GRB}(z)}{1+z}\frac{\mathrm{d}V(z)}{\mathrm{d}z}\;\mathrm{d}z\\
&\times\int_{L_{\rm lim}(z)}^{\infty}\phi(L)\;\mathrm{d}L\;,
\end{aligned}
\end{equation}
where $\Delta T\sim21\;\mathrm{yr}$ is the observational period of \emph{Swift}
that covers our sample, $\Delta\Omega$ is the angular sky coverage, $F(z)$
represents the probability of triggering a GRB and successfully measuring its
redshift, $\langle f_{\rm beam}\rangle$ is the average beaming factor (i.e.,
the fraction of the sky covered by a typical GRB jet), $(1+z)^{-1}$ accounts for
cosmological time dilation, $\mathrm{d}V(z)/\mathrm{d}z$ is the comoving volume
element, and $\phi(L)$ is the normalized GRB LF. Since all GRBs in the current
\emph{Swift} sample have $z<10$, we set $z_{\rm max}=10$. The luminosity
threshold in Equation~(\ref{eq:Number}) can be approximated using the bolometric
energy flux limit $F_{\rm lim}=3.0\times10^{-8}\,\mathrm{erg\,cm^{-2}\,s^{-1}}$,
yielding
\begin{equation}
\label{eq:Llimit}
L_{\rm lim}(z)=4\pi D_L^{2}(z) F_{\rm lim}\;.
\end{equation}

In practice, not all faint bursts with peak flux slightly above the instrument
threshold are successfully triggered. The \emph{Swift}/BAT trigger system is
complex, and its sensitivity to GRBs is difficult to model precisely
\citep{2006ApJ...644..378B}. Furthermore, not all triggered bursts have redshift
measurements; those with measured redshifts account for only about one-third of
all GRBs. Redshift measurement itself is a complicated process, depending on
several observational factors, such as optical follow-up observations and
spectral-line confirmation \citep{2003AJ....125.2865B}. In addition, the
conditions favorable for optical follow-up searches are difficult to parameterize
reliably. To avoid the complications arising from a detailed treatment of the
\emph{Swift} threshold and the low completeness of redshift measurements, we adopt
a model-independent strategy. Specifically, following \cite{2008ApJ...673L.119K},
we restrict our analysis to bright GRBs satisfying $L\ge L_{\rm lim}$ and $z<10$.
The luminosity cut is chosen to equal the threshold at the highest redshift of
the sample, i.e., $L_{\rm lim}(z=10)\simeq4.0\times10^{52}\,\mathrm{erg\,s^{-1}}$.
This cut minimizes selection effects in the GRB data. Moreover, by restricting
our analysis to bursts whose luminosities are sufficiently high to be detectable
across the entire redshift range, we ensure that the probability of triggering
a GRB and successfully measuring its redshift---denoted $F(z)$ in
Equation~(\ref{eq:Number})---can be treated as a constant $F_0$
\citep{2008ApJ...673L.119K}.\footnote{
The luminosity cut is applied to ensure completeness of the sample.
Even if the high-redshift ($z>8$) completeness were lower than assumed,
our constraints would become more conservative, since missing bursts
would reduce the observed counts and thus weaken the lower limits on $m_x$.}
Under these conditions, we obtain a subsample of 118 GRBs, delimited by
the green diagonal region in Figure~\ref{fig2}, which includes nine
high-redshift ($z\ge6$) bursts.

Because $L_{\rm lim}$ is now constant, the integral of the LF in
Equation~(\ref{eq:Number}) becomes a constant coefficient, regardless of
the specific form of $\phi(L)$. Consequently, Equation~(\ref{eq:Number}) reduces to
\begin{equation}
\label{eq:Nexp}
N_{\rm exp}=\Delta T\mathcal{K}
\int_{0}^{z_{\rm max}}\frac{\left(1+z\right)^{\alpha}\dot{\rho}_{\star}\left(z;\,m_{x}\right)}{1+z}\frac{\mathrm{d}V(z)}{\mathrm{d}z}\;\mathrm{d}z\;,
\end{equation}
where $\mathcal{K}=\Delta\Omega F_{0} \langle f_{\rm beam}\rangle \eta_{0}\int_{L_{\rm lim}}^{\infty}\phi(L)\;\mathrm{d}L/4\pi$
is the combined constant coefficient.

In this work, we use the maximum likelihood method, first introduced by
\cite{1983ApJ...269...35M}, to constrain the model parameters. The likelihood
function is defined as (e.g.,
\citealt{1998ApJ...496..752C,2006ApJ...643...81N,2009ApJ...699..603A,2012ApJ...751..108A,2019MNRAS.488.4607L,2021MNRAS.508...52L,2022ApJ...938..129L,2025MNRAS.542.1841G})
\begin{equation}
\mathcal{L}=\exp\left(-N_{\exp}\right) \prod_{i=1}^{N_{\mathrm{obs}}}\Phi\left(z_{i},\,t_{i}\right)\,,
\label{eq:likelihood}
\end{equation}
where $N_{\rm exp}$ is the expected number of GRB detections, estimated using
Equation~(\ref{eq:Nexp}), $N_{\rm obs}=118$ is the observed size of the subsample,
and $\Phi(z_{i},\,t_{i})$ denotes the differential rate of bursts per unit time
and per unit redshift, evaluated at $(z_{i},\,t_{i})$. This rate can be written as
\begin{equation}
\label{eq:diffrate}
\Phi\left(z,\,t\right)=\frac{\mathrm{d}^{2}N}{\mathrm{d}t\mathrm{d}z}=\mathcal{K}\frac{\left(1+z\right)^{\alpha}\dot{\rho}_{\star}\left(z;\,m_{x}\right)}{1+z}\frac{\mathrm{d}V(z)}{\mathrm{d}z}\,.
\end{equation}

We work with the logarithm of the likelihood, which attains its maximum at the
same parameter values as $\mathcal{L}$. The log-likelihood for the subsample is
\begin{equation}
\label{eq:chi2}
\begin{aligned}
\ln\mathcal{L}&=-N_{\exp}+\sum_{i=1}^{N_{\mathrm{obs}}}\ln\left[\Phi\left(z_{i},\,t_{i}\right)\right]\\
&\equiv-\mathcal{K}\mathcal{A}+N_{\mathrm{obs}}\ln \mathcal{K}+\mathcal{B}\;,
\end{aligned}
\end{equation}
where
\begin{equation}
\mathcal{A}\equiv \Delta T
\int_{0}^{z_{\rm max}}\frac{\left(1+z\right)^{\alpha}\dot{\rho}_{\star}\left(z;\,m_{x}\right)}{1+z}\frac{\mathrm{d}V(z)}{\mathrm{d}z}\;\mathrm{d}z
\end{equation}
and
\begin{equation}
\mathcal{B}\equiv \sum_{i=1}^{N_{\mathrm{obs}}}\ln\left[\frac{\left(1+z_i\right)^{\alpha}\dot{\rho}_{\star}\left(z_i;\,m_{x}\right)}{1+z_i}\frac{\mathrm{d}V(z_i)}{\mathrm{d}z}\right]\;.
\end{equation}
Given the form of Equation~(\ref{eq:chi2}), we can marginalize the nuisance
coefficient $\mathcal{K}$ by setting $\partial\ln\mathcal{L}/\partial\mathcal{K}=0$,
which yields the solution $\mathcal{K}=N_{\mathrm{obs}}/\mathcal{A}$. This
procedure allows us to determine the optimized value of $\mathcal{K}$ along
with the best-fit parameter combination $\{m_x,\;\alpha\}$. In practice, we use
$1/m_x$ as the free parameter for the likelihood analysis to avoid numerical
divergences when approaching the CDM limit (i.e., $m_x \rightarrow \infty$).

\begin{figure}
%\vskip-0.1in
\centerline{\includegraphics[angle=0,scale=0.4]{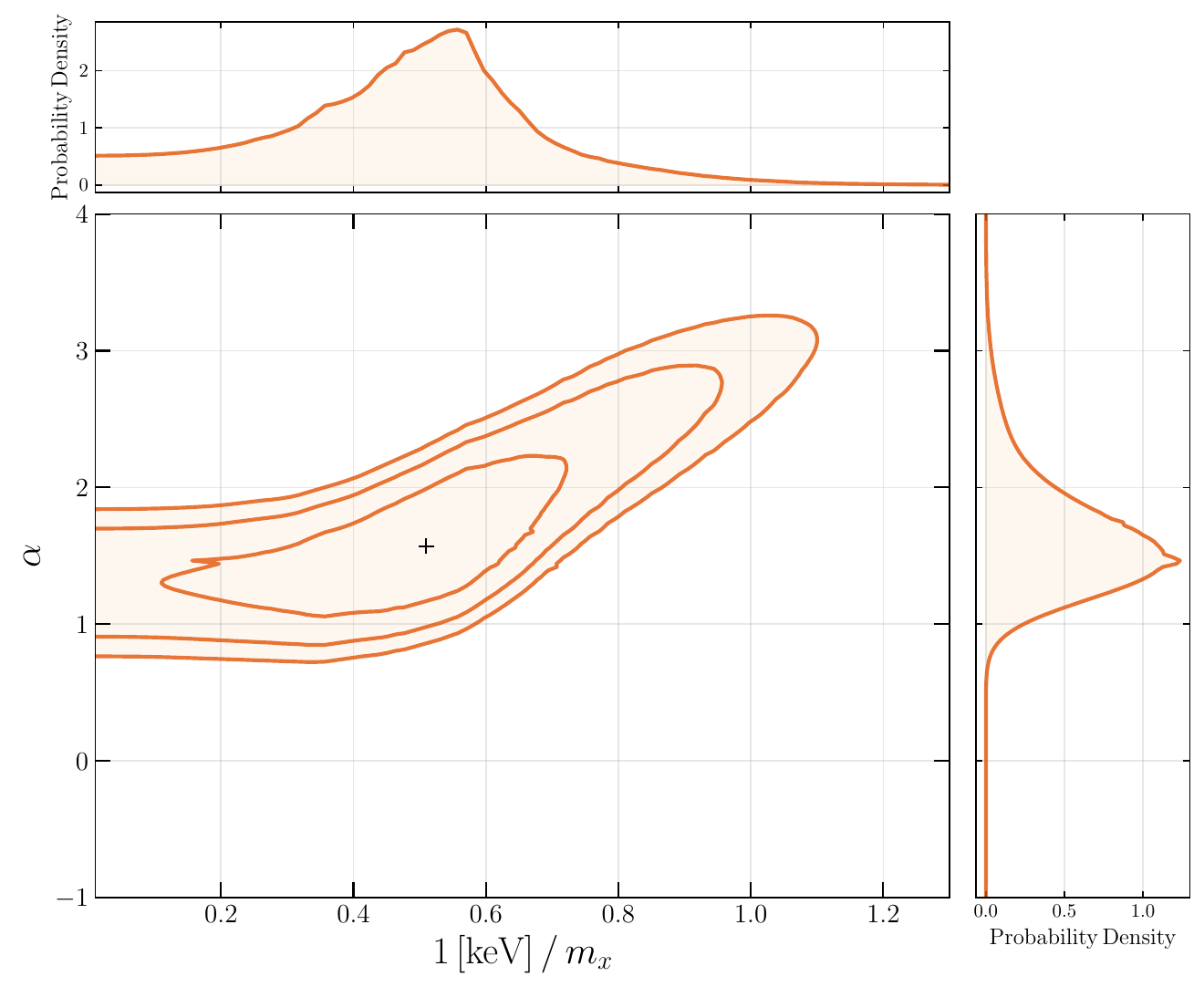}}
\caption{1D and 2D marginalized probability distributions for the inverse WDM particle
mass, $1/m_x$, and the evolutionary index $\alpha$, with 68\%, 95\%, and 99\%
confidence contours. The plus sign masks the best-fit parameter combination.}
\label{fig3}
\end{figure}

\subsection{Results}
Using the maximum likelihood analysis method described in Section~\ref{subsec:MLE},
we optimize the free parameters of our model. After marginalizing over
the nuisance coefficient $\mathcal{K}$, we obtain the one-dimensional (1D) and
two-dimensional (2D) marginalized probability distributions for the inverse
WDM particle mass, $1/m_x$, and the evolutionary index $\alpha$. The 68\%, 95\%,
and 99\% confidence regions, derived from our subsample of 118 GRBs with $z<10$ and
$L\ge 4.0\times10^{52}\,\mathrm{erg\,s^{-1}}$, are presented in Figure~\ref{fig3}.
These contours reveal a strong degeneracy between $1/m_x$ and $\alpha$, and
yield the following 95\% confidence intervals: $1\,\mathrm{keV}/m_x\lesssim0.81$
(upper limit) and $\alpha=1.57^{+1.14}_{-0.57}$. The quoted error bars correspond to
the 2.5th and 97.5th percentiles of the 1D posterior distribution of $\alpha$.
Consequently, the no-evolution case ($\alpha=0$), corresponding to the assumption
that the GRB formation rate strictly traces the SFR, is excluded at the $5\sigma$
level. In addition, WDM particle masses below $1.23\,\mathrm{keV}$ are excluded
at the 95\% CL.

\begin{figure}
%\vskip-0.1in
\centerline{\includegraphics[angle=0,scale=0.55]{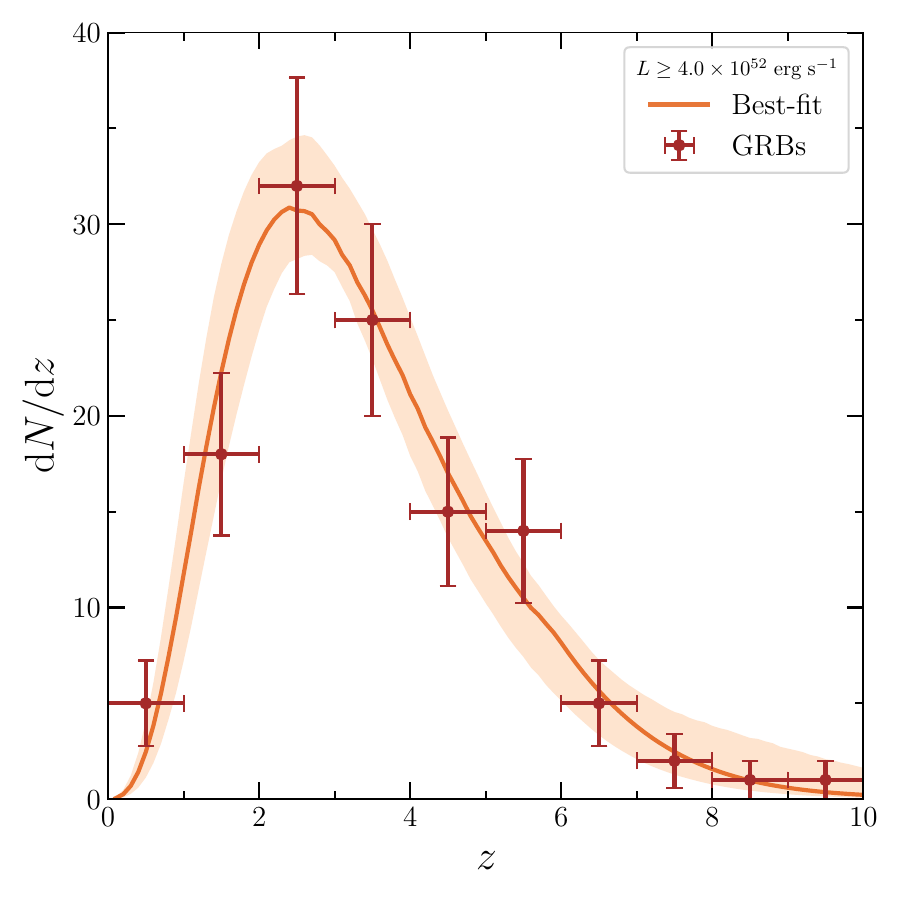}}
\caption{Redshift distribution of 118 \emph{Swift} GRBs with
$L\ge 4.0\times10^{52}\,\mathrm{erg\,s^{-1}}$. Data points show the observed number
of bursts in each redshift bin, with Poisson error bars.
The solid line represents the redshift distribution predicted by the best-fit model,
and the shaded region indicates the 95\% confidence interval of the best-fit model.}
\label{fig4}
\end{figure}

Figure~\ref{fig4} shows the redshift distribution of the 118 GRBs in our subsample,
which satisfy $L\ge 4.0\times10^{52}\,\mathrm{erg\,s^{-1}}$. The distribution
predicted by the best-fit model (solid line) is in good agreement with the observed
redshift distribution of the subsample.

\begin{figure}
%\vskip-0.1in
\centerline{\includegraphics[angle=0,scale=0.55]{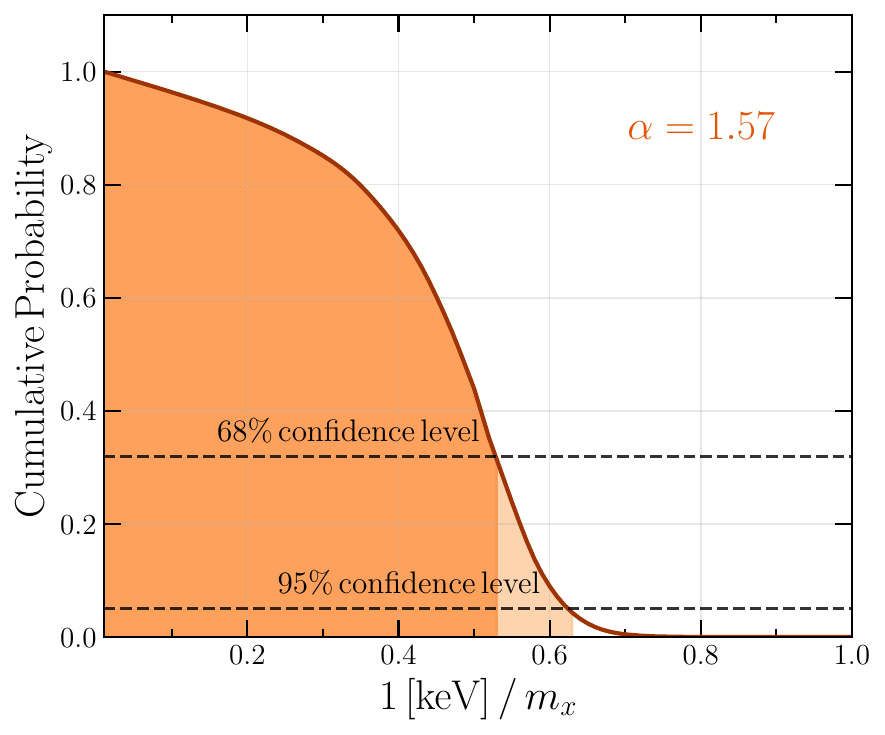}}
\caption{Cumulative posterior distribution of the inverse WDM particle mass,
$1/m_x$, obtained with a prior $\alpha=1.57$. The dark and light shaded regions mark
the survival values at the 68\% and 95\% CLs, respectively.}
\label{fig5}
\end{figure}

If a prior of $\alpha=1.57$ (the best-fit value) is adopted, the resulting
cumulative probability distribution for $1/m_x$ is shown in Figure~\ref{fig5}.
The 95\% CL upper limit on $1/m_x$ improves to $1\,\mathrm{keV}/m_x\lesssim0.63$.
With this prior, models with $m_x \lesssim 1.59\,\mathrm{keV}$ are therefore
ruled out at the 95\% CL.

As shown in Figure~\ref{fig1}, higher-redshift data are more sensitive to the WDM
particle mass, but are also more susceptible to the complex astrophysics of the
Pop III stellar population. To quantify the contribution of the $z>6$ data to
our constraints, we perform a parallel comparative analysis using a subsample
of 109 GRBs with $z<6$ and $L\ge 4.0\times10^{52}\,\mathrm{erg\,s^{-1}}$.
We find that excluding the nine high-redshift ($z>6$) GRBs yields substantially
weaker constraints on both $m_x$ and $\alpha$, giving $m_x \gtrsim 1.03\,\mathrm{keV}$
and $\alpha=1.63^{+1.45}_{-0.54}$ at the 95\% CL. This indicates that the inclusion
of high-redshift bursts significantly enhances the constraining power, and that
future observations with an enlarged sample of such bursts would further tighten
the limits on $m_x$.

\section{Conclusions and Discussion}
\label{sec:summary}
Although the CDM paradigm has been remarkably successful in explaining
the large-scale structure of the Universe, it faces persistent challenges
on small scales. An alternative WDM paradigm, with particle masses in the
keV range, can alleviate these problems by strongly suppressing small-scale
structure formation. The high-redshift Universe provides a powerful testbed
for such WDM models, as the mere presence of collapsed structures at early
epochs can set strong lower limits on the WDM particle mass $m_x$. Thanks to
their extreme brightness and detectability deep into the early Universe,
GRBs offer a promising means for constraining WDM models.

In this work, we adopt a self-consistent method to model the cosmic SFR
within the WDM framework. The well-established connection between long GRBs
and the core collapse of massive stars allows us to derive the expected redshift
distribution of long GRBs directly from this SFR model as a function of the
WDM particle mass $m_x$. Thus, $m_x$ can be constrained by directly comparing
the observed redshift distribution of GRBs with the model prediction.
To perform a more robust statistical analysis, we update and expand the sample
of \emph{Swift} GRBs with measured redshifts accumulated over the past two decades.
We conservatively account for astrophysical uncertainties by assuming that
the comoving GRB formation rate is proportional to the cosmic SFR, with
an additional redshift evolution parameterized as $(1+z)^\alpha$. To reduce
complications arising from instrumental selection effects, we adopt a
model-independent strategy: we select only luminous GRBs above a fixed
luminosity threshold. The high statistical quality of the latest GRB sample
enables us to apply such a luminosity cut to fairly compare GRBs across
the entire redshift range, thereby circumventing the need to model the unknown GRB LF.

For each model parameter combination ($1/m_x$, $\alpha$), we compute
the expected redshift distribution of GRBs. Applying a maximum-likelihood
estimator to the subsample of 118 \emph{Swift} GRBs with $z<10$ and
$L\ge 4.0\times10^{52}\,\mathrm{erg\,s^{-1}}$, we simultaneously
constrain $1/m_x$ and $\alpha$.
Our results indicate that $1\,\mathrm{keV}/m_x\lesssim0.81$ and
$\alpha=1.57^{+1.14}_{-0.57}$ at the 95\% CL. Consequently, the no-evolution scenario
($\alpha=0$), in which the GRB formation rate directly follows the SFR without
additional redshift evolution, is excluded at the $5\sigma$ level. In addition,
the WDM particle mass is constrained to $m_x \gtrsim 1.23\,\mathrm{keV}$
at the 95\% CL. When the best-fit value $\alpha=1.57$ is adopted as a prior,
this lower limit tightens to $m_x \gtrsim 1.59\,\mathrm{keV}$ at the same CL.
Our constraints on WDM models are strong and robust, demonstrating that
GRBs serve as a powerful probe of the early Universe. Furthermore, if
the mechanism governing the relationship between the GRB rate and the SFR
were better understood, GRBs could impose even tighter constraints on WDM models.

In our construction of the cosmic star formation history, we adopt global,
redshift-independent star-formation timescales ($\tau_1 =3.5\,\mathrm{Gyr}$
for Pop II/I stars and $\tau_2 =0.1\,\mathrm{Gyr}$ for Pop III stars). While
this simplification does not account for the modified accretion rates and
halo concentration profiles expected in WDM cosmologies, we argue that
it does not significantly affect our conclusions. The GRB sample used
in this work is limited to $z<10$, a redshift range over which \citet{2019MNRAS.486.3617L}
have shown, using cosmological zoom-in simulations, that the differences
between CDM and WDM in globally averaged observables, such as the SFR density
and mean metallicity, already diminish to within a factor of $\sim2.5$ by $z\sim8.5$.
Thus, a simple scaling prescription is adequate for the present analysis. Nonetheless,
we recognize that WDM may alter the spatial distribution of metal enrichment,
which could in turn affect the properties of GRB host galaxies. Our constraints
should therefore be interpreted with this caveat in mind, and future work that
explicitly incorporates WDM-specific accretion physics would further strengthen
the robustness of the limits.

\section*{Acknowledgements}
We are grateful to Dr. Yan-Kun Qu for his assistance in compiling
the GRB sample with known redshifts, and to Dr. Wei-Wei Tan for
fruitful discussions. This work is supported by the National
Key R\&D Program of China (grant Nos. 2024YFA1611704 and 2024YFA1611700),
the Strategic Priority Research Program of the Chinese Academy of Sciences
(grant No. XDB0550400), the National Natural Science Foundation of China
(grant Nos. 12422307 and 12373053), and China Postdoctoral
Science Foundation (grant No. 2025M783235).
We are also grateful to the anonymous referee for helpful comments.

%%%%%%%%%%%%%%%%%%%%%%%%%%%%%%%%%%%%%%%%%%%%%%%%%%
\section*{Data Availability}
The data underlying this article will be shared on reasonable request
to the corresponding author.



%\bibliographystyle{mnras}
%\bibliography{references} % if your bibtex file is called example.bib

%\begin{thebibliography}{99}
%\end{thebibliography}

%%%%%%%%%%%%%%%%%%%%%%%%%%%%%%%%%%%%%%%%%%%%%%%%%%

%%%%%%%%%%%%%%%%% APPENDICES %%%%%%%%%%%%%%%%%%%%%
%\clearpage
\appendix

\section{Additional table}

%\nopagebreak[4]
\begin{table*} %[H]
\centering
\renewcommand{\arraystretch}{1.1}
\caption{The 73 newly added GRBs with known redshifts.$^{a}$}
\label{table:A1}
\resizebox{\textwidth}{!}{
\begin{tabular}{lccc|lccc}
\hline\hline
GRB & Peak flux$^{b}$ & \emph{z} & ${\log_{10}L}^{c}$ & GRB & Peak flux$^{b}$ & \emph{z} & ${\log_{10}L}^{c}$ \\
    & $(\mathrm{photons\,cm^{-2}\,s^{-1}})$ &  & $(\mathrm{erg\,s^{-1}})$ &     & $(\mathrm{photons\,cm^{-2}\,s^{-1}})$ &  & $(\mathrm{erg\,s^{-1}})$ \\
\hline
191221B	&$	4.80	\pm	0.70	$&$	1.19	$&$	52.09	\pm	0.06	$&	230506C	&$	2.40	\pm	0.50	$&$	3.7	$&$	52.94	\pm	0.09	$\\
200205B	&$	2.00	\pm	0.20	$&$	1.465	$&$	51.92	\pm	0.04	$&	230818A	&$	6.70	\pm	0.50	$&$	2.42	$&$	52.96	\pm	0.03	$\\
200829A	&$	98.70	\pm	1.60	$&$	1.25	$&$	53.45	\pm	0.01	$&	231111A	&$	2.70	\pm	0.30	$&$	1.179	$&$	51.83	\pm	0.05	$\\
201014A	&$	0.30	\pm	0.20	$&$	4.56	$&$	52.24	\pm	0.29	$&	231118A	&$	9.70	\pm	0.50	$&$	0.8304	$&$	52.02	\pm	0.02	$\\
201015A	&$	1.80	\pm	0.40	$&$	0.426	$&$	50.60	\pm	0.10	$&	231210B	&$	3.90	\pm	0.70	$&$	3.13	$&$	52.98	\pm	0.08	$\\
201020A	&$	2.10	\pm	0.20	$&$	2.903	$&$	52.64	\pm	0.04	$&	231215A	&$	9.60	\pm	1.40	$&$	2.305	$&$	53.07	\pm	0.06	$\\
201021C	&$	1.00	\pm	0.20	$&$	1.07	$&$	51.29	\pm	0.09	$&	240205B	&$	47.80	\pm	1.00	$&$	0.824	$&$	52.70	\pm	0.01	$\\
201024A	&$	3.70	\pm	0.60	$&$	0.999	$&$	51.79	\pm	0.07	$&	240218A	&$	3.00	\pm	0.30	$&$	6.782	$&$	53.61	\pm	0.04	$\\
201104B	&$	4.10	\pm	0.30	$&$	1.954	$&$	52.53	\pm	0.03	$&	240414A	&$	1.50	\pm	0.20	$&$	1.833	$&$	52.03	\pm	0.06	$\\
201216C	&$	18.00	\pm	1.10	$&$	1.1	$&$	52.58	\pm	0.03	$&	240419A	&$	0.60	\pm	0.10	$&$	5.178	$&$	52.66	\pm	0.07	$\\
201221A	&$	1.00	\pm	0.30	$&$	5.7	$&$	52.97	\pm	0.13	$&	240529A	&$	7.70	\pm	0.60	$&$	2.695	$&$	53.13	\pm	0.03	$\\
210210A	&$	7.00	\pm	0.50	$&$	0.715	$&$	51.72	\pm	0.03	$&	240809A	&$	12.80	\pm	0.40	$&$	1.495	$&$	52.75	\pm	0.01	$\\
210222B	&$	1.40	\pm	0.30	$&$	2.198	$&$	52.18	\pm	0.09	$&	240825A	&$	100.00	\pm	1.70	$&$	0.659	$&$	52.79	\pm	0.01	$\\
210321A	&$	1.70	\pm	0.40	$&$	1.487	$&$	51.87	\pm	0.10	$&	240912A	&$	15.80	\pm	0.50	$&$	1.234	$&$	52.64	\pm	0.01	$\\
210411C	&$	4.80	\pm	0.30	$&$	2.826	$&$	52.97	\pm	0.03	$&	241010A	&$	4.30	\pm	0.20	$&$	0.977	$&$	51.83	\pm	0.02	$\\
210420B	&$	0.50	\pm	0.20	$&$	1.4	$&$	51.27	\pm	0.17	$&	241026A	&$	4.00	\pm	0.50	$&$	2.79	$&$	52.88	\pm	0.05	$\\
210504A	&$	0.90	\pm	0.20	$&$	2.077	$&$	51.93	\pm	0.10	$&	241030A	&$	11.80	\pm	0.30	$&$	1.411	$&$	52.65	\pm	0.01	$\\
210517A	&$	1.50	\pm	0.20	$&$	2.486	$&$	52.34	\pm	0.06	$&	241030B	&$	3.20	\pm	0.60	$&$	2.82	$&$	52.79	\pm	0.08	$\\
210610A	&$	2.40	\pm	0.40	$&$	3.54	$&$	52.89	\pm	0.07	$&	250101A	&$	0.80	\pm	0.20	$&$	2.49	$&$	52.07	\pm	0.11	$\\
210610B	&$	13.50	\pm	0.70	$&$	1.13	$&$	52.48	\pm	0.02	$&	250108B	&$	0.60	\pm	0.20	$&$	2.197	$&$	51.82	\pm	0.14	$\\
210619B	&$	115.00	\pm	2.20	$&$	1.937	$&$	53.97	\pm	0.01	$&	250114A	&$	1.50	\pm	0.40	$&$	4.732	$&$	52.97	\pm	0.12	$\\
210722A	&$	2.20	\pm	0.50	$&$	1.145	$&$	51.71	\pm	0.10	$&	250129A	&$	1.40	\pm	0.20	$&$	2.151	$&$	52.16	\pm	0.06	$\\
210731A	&$	1.60	\pm	0.30	$&$	1.2525	$&$	51.66	\pm	0.08	$&	250225B	&$	10.40	\pm	0.50	$&$	0.95	$&$	52.19	\pm	0.02	$\\
210822A	&$	27.70	\pm	0.70	$&$	1.736	$&$	53.24	\pm	0.01	$&	250424A	&$	47.00	\pm	1.30	$&$	0.31	$&$	51.69	\pm	0.01	$\\
210905A$^{d}$	&$	3.83^{+0.73}_{-0.54}\times10^{-7}$&$	6.318	$&$	53.27^{+0.08}_{-0.06}$&	250430A	&$	2.30	\pm	0.20	$&$	0.767	$&$	51.31	 \pm	0.04	$\\
211024B	&$	0.90	\pm	0.20	$&$	1.1137	$&$	51.29	\pm	0.10	$&	250617B	&$	2.90	\pm	0.80	$&$	1.396	$&$	52.03	\pm	0.12	$\\
211207A	&$	0.90	\pm	0.20	$&$	2.272	$&$	52.03	\pm	0.10	$&	250702F	&$	5.10	\pm	0.30	$&$	1.52	$&$	52.37	\pm	0.03	$\\
220101A	&$	7.30	\pm	0.30	$&$	4.61	$&$	53.63	\pm	0.02	$&	250725A	&$	12.50	\pm	0.60	$&$	5.26	$&$	53.99	\pm	0.02	$\\
220117A	&$	1.90	\pm	0.40	$&$	4.961	$&$	53.12	\pm	0.09	$&	250807B	&$	0.90	\pm	0.30	$&$	1.522	$&$	51.61	\pm	0.14	$\\
220521A	&$	4.70	\pm	0.40	$&$	5.6	    $&$	53.63	\pm	0.04	$&	250920C	&$	15.90	\pm	1.00	$&$	1.4	$&$	52.78	\pm	0.03	$\\
221009A	&$	1.90	\pm	0.30	$&$	0.151	$&$	49.60	\pm	0.07	$&	250925A	&$	1.20	\pm	0.20	$&$	3.899	$&$	52.69	\pm	0.07	$\\
221110A	&$	1.10	\pm	0.20	$&$	4.06	$&$	52.69	\pm	0.08	$&	251001B	&$	0.80	\pm	0.10	$&$	2.162	$&$	51.92	\pm	0.05	$\\
221226B	&$	2.60	\pm	0.30	$&$	2.694	$&$	52.66	\pm	0.05	$&	251017A	&$	1.30	\pm	0.20	$&$	4.34	$&$	52.82	\pm	0.07	$\\
230116D	&$	0.60	\pm	0.20	$&$	3.81	$&$	52.36	\pm	0.14	$&	251126A	&$	1.30	\pm	0.20	$&$	3.52	$&$	52.62	\pm	0.07	$\\
230325A	&$	3.30	\pm	0.40	$&$	1.664	$&$	52.27	\pm	0.05	$&	251205A	&$	0.50	\pm	0.20	$&$	1.1	$&$	51.02	\pm	0.17	$\\
230328B	&$	8.40	\pm	0.50	$&$	0.09	$&$	49.77	\pm	0.03	$&	260101A	&$	4.60	\pm	0.30	$&$	2.623	$&$	52.88	\pm	0.03	$\\
230414B	&$	0.60	\pm	0.20	$&$	3.568	$&$	52.30	\pm	0.14	$&	     	&	  	                 &      	 & 	                     \\
\hline\hline
\end{tabular}
}
\begin{description}
  \item[\emph{Note.}] {$^{a}$All GRB information is taken from \url{https://swift.gsfc.nasa.gov/archive/grb_table/}.
  $^{b}$1-second peak photon flux in the 15--150 keV {\it Swift}/BAT energy band
(with 90\% CL uncertainty). $^{c}$Isotropic peak luminosity calculated in the
1--$10^4$ keV rest-frame energy range, with the 90\% CL uncertainty propagated
solely from the peak flux error. $^{d}$For GRB 210905A, we use the 1-second peak energy flux
(in units of $\mathrm{erg\,cm^{-2}\,s^{-1}}$) in the 15--1500 keV energy range and
the isotropic peak luminosity reported by \cite{2022A&A...665A.125R}.}
\end{description}
\end{table*}

%%%%%%%%%%%%%%%%%%%%%%%%%%%%%%%%%%%%%%%%%%%%%%%%%%

% Don't change these lines
\bsp	% typesetting comment
\label{lastpage}
\end{document}